\documentclass[]{jfm}

\usepackage{graphicx}
\usepackage{newtxtext}
\usepackage{newtxmath}
\usepackage{natbib}
\usepackage{hyperref}
\hypersetup{
    colorlinks = true,
    urlcolor   = blue,
    citecolor  = black,
}
\usepackage{booktabs}

\usepackage{bm}

\usepackage[svgnames]{xcolor}
\hypersetup{colorlinks=true, citecolor=NavyBlue, urlcolor=NavyBlue, linkcolor=NavyBlue}

\title{Bistability of the large-scale dynamics in quasi-two-dimensional turbulence}

\author{Xander M. de Wit\aff{1,2},
  Adrian van Kan\aff{1,3} \and
  Alexandros Alexakis\aff{1}
  \corresp{\email{alexandros.alexakis@phys.ens.fr}}}

\affiliation{
\aff{1}Laboratoire de Physique de l’Ecole Normale Supérieure, ENS, Université PSL, CNRS,
Sorbonne Université, Université Paris-Diderot, Sorbonne Paris Cité, Paris, France\\
\aff{2}Fluids and Flows group, Department of Applied Physics and J. M. Burgers Centre for Fluid Dynamics, Eindhoven University of Technology, P.O. Box 513, 5600 MB Eindhoven, Netherlands\\
\aff{3}Department of Physics, University of California, Berkeley, CA 94720, USA}

\begin{document}
\maketitle

\begin{abstract}
In many geophysical and astrophysical flows, suppression of fluctuations along one direction of the flow drives a quasi-2D upscale flux of kinetic energy, leading to the formation of strong vortex condensates at the largest scales. Recent studies have shown that the transition towards this condensate state is hysteretic, giving rise to a limited bistable range in which both the condensate state as well as the regular 3D state can exist at the same parameter values. In this work, we use direct numerical simulations of thin-layer flow to investigate whether this bistable range survives as the domain size and turbulence intensity are increased. By studying the time scales at which rare transitions occur from one state into the other, we find that the bistable range grows as the box size and/or Reynolds number $\Rey$ are increased, showing that the bistability is neither a finite-size nor a finite-$\Rey$ effect. We furthermore predict a crossover from a bimodal regime at low box size, low $\Rey$ to a regime of pure hysteresis at high box size, high $\Rey$, in which any transition from one state to the other is prohibited at any finite time scale.
\end{abstract}


\section{Introduction} \label{sec:intro} 
\begin{figure}
    \centering
    \includegraphics[width=0.7\linewidth]{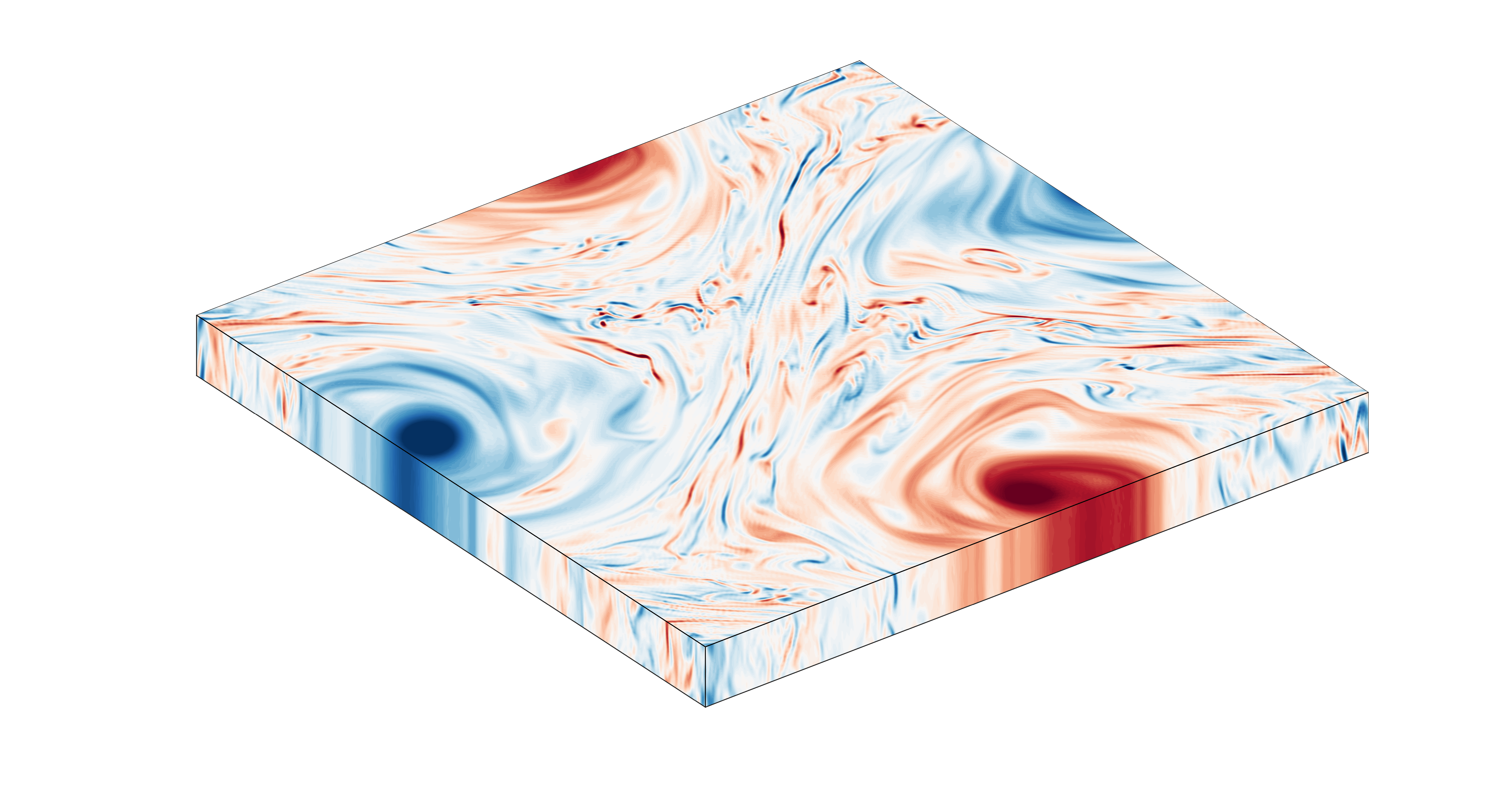}
    \caption{A vortex condensate in thin-layer flow, visualised through a snapshot of vertical vorticity.}
    \label{fig:condensate_snap}
\end{figure}

Ever since the seminal works of \citet{Batchelor1969} and \citet{Kraichnan1967}, it has been known that in 2D turbulence, contrary to what is observed in 3D turbulence, kinetic energy cascades inversely, from the smaller scales at which it is injected to ever larger and larger scales. While the forward cascade that is observed in 3D turbulence is always arrested once it arrives at the scales at which viscosity becomes effective in dissipating the kinetic energy, such a stopping mechanism does not always exist at the large scales to saturate the inverse cascade. In that case, kinetic energy piles up at the largest available length scale of the flow system into what is referred to as a condensate. This condensate typically manifests as a strong vortex structure at the system size, also known as the Large-Scale Vortex, see figure~\ref{fig:condensate_snap}.

Even in 3D flow systems, quasi-2D dynamics can be observed if fluctuations in one direction are strongly suppressed, allowing an inverse cascade to develop \citep{Alexakis2018}. In forced rotating turbulence \citep{Biferale2016,Mininni2009,Smith1996,Smith1999}, rotating convection \citep{Favier2014,Guervilly2014,Julien2012,Rubio2014} or rotating stratified turbulence \citep{Pouquet2013,Marino2013,Marino2014,VanKan2020}, such quasi-2D dynamics develops as a consequence of the Coriolis force, which inhibits the transfer of energy to eddies varying along the axis of rotation. Alternatively, such suppression could occur through for example magnetic forces \citep{Alexakis2011,Baker2018,Favier2010,Reddy2014} or plainly through geometric confinement as observed in thin-layer flow \citep{Celani2010,Benavides2017,Musacchio2017,Musacchio2019}. This type of constrained dynamics is of eminent importance to many geo- and astrophysical flow settings, where (a combination of) the aforementioned mechanisms renders the flow quasi-2D. Examples can be found in our oceans \citep{King2015,Scott2005}, in the atmosphere \citep{Byrne2013,Nastrom1984} and on gas-giant planets such as Jupiter and Saturn \citep{Heimpel2007,Heimpel2016,Stellmach2016}.

This work focuses on the transition towards the condensate state of such quasi-2D systems. Remarkably, in spite of the inherently widely different nature of the considered flow systems, recent studies have revealed that all across forced rotating turbulence \citep{Alexakis2015,Seshasayanan2018,Yokoyama2017}, thin-layer turbulence \citep{VanKan2019} and even the natural system of rotating convection \citep{Favier2019,DeWit2022}, the transition into the condensate is discontinuous and shows hysteresis. This gives rise to a limited bistable range in which both the quasi-2D condensate state and the 3D flow state can exist at the same parameters.

Since it is now known that this bistability can also survive in natural forcing conditions \citep{Favier2019,DeWit2022} and the condensate can also form, albeit at more extreme parameters, between realistic no-slip walls \citep{Guzman2020}, we aim to investigate whether the bistable range in the condensate transition could also survive under parameter conditions that are relevant to geo- and astrophysical flows. Motivated by the remarkable similarities in the condensate transition across the different flow systems, we focus on the conceptually and computationally most basic system of forced thin-layer turbulence. Specifically, we are interested in the dependence on the system size and the strength of turbulent forcing, quantified through the injection-scale Reynolds number $\Rey$, in order to investigate whether the bistable range of the condensate transition shrinks or grows as system size and $\Rey$ are increased. We focus on very moderate values of the system size and $\Rey$ in order to be able to gather computationally demanding statistics about the bistable range and the rare transitions into and out of the condensate state. By identifying the system size and $\Rey$ dependence, we can then obtain a first clue as to whether the bistable behaviour could possibly be observed in the limits of large system size and large $\Rey$ that are relevant to the real-world natural geophysical and astrophysical flow settings.

\section{Numerical approach}\label{sec:numerical_approach}  
In order to study thin-layer turbulence, we consider the idealised case of forced incompressible 3D flow in a triply periodic box of dimensions $L \times L \times H$, where the vertical direction is thin $H\ll L$. The flow system is identical to that described in \citet{VanKan2019}. We consider a Cartesian coordinate system $(x,y,z)$ with unit vectors $(\boldsymbol{e}_x,\boldsymbol{e}_y,\boldsymbol{e}_z)$, where the thin vertical direction is chosen along $\boldsymbol{e}_z$. The flow $\boldsymbol{u}(\boldsymbol{x},t)$ is then governed by the incompressible forced Navier-Stokes equations
\begin{subequations}\begin{align}
    \frac{\partial\boldsymbol{u}}{\partial t}+(\boldsymbol{u}\cdot\boldsymbol{\nabla})\boldsymbol{u}&=-\boldsymbol{\nabla} P+\nu\nabla^2\boldsymbol{u}+\boldsymbol{f},\label{eq:gov1}\\
    \boldsymbol{\nabla}\cdot\boldsymbol{u}&=0,\label{eq:gov2}
\end{align}\end{subequations}
where $P(\boldsymbol{x},t)$ denotes the pressure divided by the constant density and $\nu$ represents the kinematic viscosity of the fluid. We consider a stochastic forcing $\boldsymbol{f}(\boldsymbol{x},t)$ that is vertically invariant ($\partial\boldsymbol{f}/\partial z=0$) and acts exclusively in the $(\boldsymbol{e}_x,\boldsymbol{e}_y)$ plane, i.e. in the two-dimensional two-component (2D2C) manifold. Furthermore, the forcing is divergence-free and acts only sharply on wavenumber $k_f\equiv2\pi/\ell$ (specifically, exclusively the modes $(k_x,k_y)=(\pm k_f,0) \textrm{ and } (0,\pm k_f)$ are forced), where its random phase is white noise (delta-correlated) in time. This results in a fixed mean injection rate $\langle \boldsymbol{u} \cdot \boldsymbol{f} \rangle = \epsilon$ that is solely prescribed by the forcing amplitude \citep{Novikov1965}. Here, $\langle \cdot \rangle$ is used to represent the ensemble average. The choice of forcing is motivated by simplicity and comparability with previous studies. In general, one may consider various 3D forcing functions \citep{Poujol2020}.

The input parameters of the thin-layer flow system are combined to give three dimensionless numbers. We define an injection-scale Reynolds number $\Rey\equiv(\epsilon \ell^4)^{1/3}/\nu$, the ratio between the forcing scale and the domain height $Q\equiv \ell/H$ and the ratio between the forcing scale and the width of the domain $K\equiv \ell/L$. Finally, we define a forcing time scale $\tau_f\equiv(\ell^2/\epsilon)^{1/3}$ and energy scale $E_f\equiv(\epsilon\ell)^{2/3}$ that are used to non-dimensionalise the different temporal and energetic quantities reported in this work, respectively.

Equations \eqref{eq:gov1}-\eqref{eq:gov2} are solved numerically in the triply periodic domain using a pseudo-spectral code that is an adapted version of the Geophysical High-Order Suite for Turbulence (\textsc{Ghost}) as introduced by \citet{Mininni2011}, employing 2/3-dealiasing. In order to investigate the dependence of the condensate transition and its bistable range on $\Rey$ and the box size, we take the results in \citet{VanKan2019a} as a starting point and extend them to smaller and larger $\Rey$ and $K$. For each value of $K,\Rey$ we vary the thinness of the fluid layer $Q$ as the principal control parameter in close vicinity to the condensate transition. An overview of the full set of input parameters that are considered in this work is provided in table~\ref{tab:input}.

Resolutions $N_x\times N_y\times N_z$ are chosen such that we maintain the same ratio between the grid spacing $L/N_{x,y}$ and Kolmogorov length $\eta=(\nu^3/\epsilon)^{1/4}$ as used in \citet{VanKan2019a} of $L/N_{x,y}\approx3.2\eta$ in the horizontal directions, and we resolve finer than that in the vertical direction. For the vertical, we ensure that we keep 16 grid cells in order to maintain sufficient degrees of freedom in the thin direction.

\begin{table}
    \renewcommand\arraystretch{1.1}
    \centering
    \caption{The different series of input parameters used in this work for varying box size (left) and varying $\Rey$ (right).}\label{tab:input}
    \begin{tabular}{cccc}
        \toprule
        $1/K$&$\Rey$&$Q$&$N_x\times N_y\times N_z$\\
        \midrule
        6&192&$[1.53:1.70]$&$96\times96\times16$\\
        7&192&$[1.53:1.69]$&$112\times112\times16$\\
        8&192&$[1.44:1.81]$&$128\times128\times16$\\
        9&192&$[1.44:1.73]$&$144\times144\times16$\\
        \bottomrule
    \end{tabular}
    \qquad
    \begin{tabular}{cccc}
        \toprule
        $1/K$&$\Rey$&$Q$&$N_x\times N_y\times N_z$\\
        \midrule
        8&76&$[1.00:1.13]$&$64\times64\times16$\\
        8&131&$[1.29:1.48]$&$96\times96\times16$\\
        8&192&$[1.44:1.81]$&$128\times128\times16$\\
        8&329&$[1.74:2.05]$&$192\times192\times16$\\
        \bottomrule
    \end{tabular}
\end{table}

For each unique set of input parameters ($Q,K,\Rey$), at least 40 independent runs are carried out by using a different random seed for the stochastic forcing in order to obtain statistics about the rare transitions from one state into the other. The combination of this need for a multitude of independent runs with the fact that the large-scale dynamics is very slow compared to the background smaller-scale 3D dynamics renders only the rather moderate parameters that are considered here computationally accessible. The simulations in this work comprise more than $\sim$1.0 million CPU hours.

The main diagnostic quantity that we use here to probe the strength of the condensate is the 2D large-scale energy $E_\textrm{ls}$, defined from the Fourier components $\hat{\boldsymbol{u}}(\boldsymbol{k})$ of the flow as
\begin{equation}
    E_\textrm{ls}=\frac{1}{2}\sum_{\substack{\boldsymbol{k}\\|\boldsymbol{k}|\leq k_\textrm{max}\\k_z=0}}\left(|\hat{\boldsymbol{u}}(\boldsymbol{k})\cdot\boldsymbol{e}_x|^2+|\hat{\boldsymbol{u}}(\boldsymbol{k})\cdot\boldsymbol{e}_y|^2\right),
\end{equation}
where the cut-off wavenumber $k_\textrm{max}=\sqrt{2}(2\pi/L)$.

\section{Bistability, bimodality and rare transitions}\label{sec:bistability} %
Bistability and bimodality are often met in dynamical systems. Here we refer to `bistability' as the presence of two independent stable attractors, coexisting for the same value of parameters, whilst by `bimodality' we refer to the case in which two attractors are linked by some trajectories such that, when followed, the system jumps from one attractor to the other and \textit{vice versa}. In the former case, a hysteresis loop can exist when one of the parameters of the system is continuously varied. In an inherently fluctuating dynamical system like the one at hand, however, the precise extent of the bistable range in the hysteresis loop can be ambiguous as rare sudden transitions from one hysteretic branch into the other may exist at very long time scales near both ends of the hysteresis loop. This raises the question as to how we can unambiguously define the precise extent of the bistable range in the hysteretic transition.

Earlier works that studied the bistable range in the quasi-2D condensate transition would typically simulate up to a certain time scale that is constrained by computational limits and call a state `stable' if no further transitions are observed \citep{Favier2019,VanKan2019,DeWit2022,Yokoyama2017}. However, in view of the rare transitions, this amounts to an in principle arbitrary cut-off at a certain time scale, neglecting any possible transitions occurring at larger time scales. We refer to this as \textit{finite-time hysteresis}.

However, in the case of \textit{pure hysteresis}, in the strict sense, all transitions from one state into the other are prohibited at any finite time scale, such that the system is absolutely bistable. This would require the time scales of the rare transitions to diverge at a certain asymptote. The existence of such asymptotes is the principal assumption in this work. As we will show in section~\ref{sec:results_discussion}, we can define such asymptotes based on the scaling of the time scales of the rare transitions that we observe, allowing us to study how these asymptotes shift as we vary the box size and $\Rey$ in order to get a completely time-scale-independent and unambiguous method for comparing our results at these different parameters.

To analyse the rare transitions between the two states, we separately consider build-up events from the 3D state into the condensate state and decay events \textit{vice versa}. The build-up events are studied by initialising the simulations with a tiny perturbation onto a state of no flow and continuing until the condensate state is reached. For the decay events, we initialise the simulation with a snapshot from a condensate state at higher $Q$ and we continue the run until the condensate has decayed and the 3D state is obtained.

\begin{figure}
    \centering
    \includegraphics[width=\linewidth]{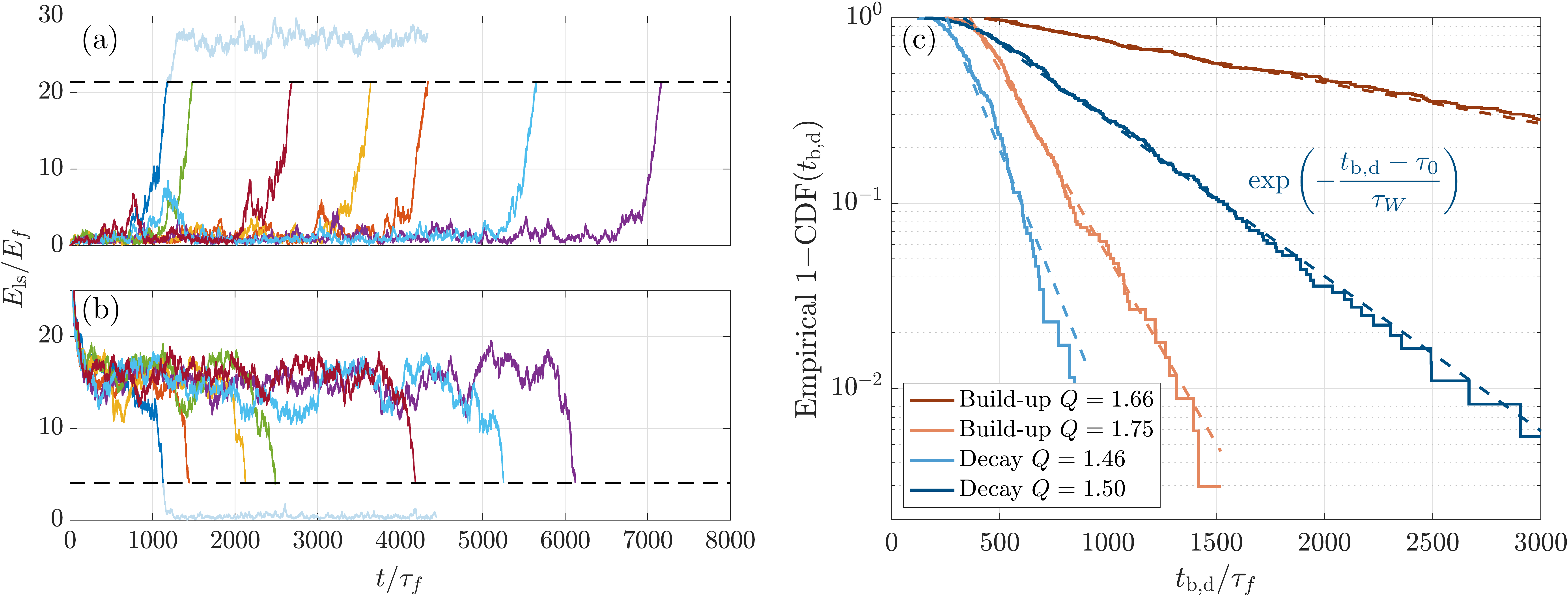}
    \caption{(a,b) Examples of time series of different realisations for (a) build-up at $Q=1.67$ and (b) decay at $Q=1.49$ (for $1/K=9$ and $\Rey=192$). Horizontal dashed lines represent the threshold energy at which the build-up time $t_b$ or decay time $t_d$ is defined and the simulation is terminated. The first blue run is continued for demonstration purposes. (c) Examples of distributions of $t_{b,d}$, shown through the empirical CDF (at $1/K=8$ and $\Rey=192$). Dashed lines represent fits of the exponential distribution.}
    \label{fig:transitions}
\end{figure}

Figures~\ref{fig:transitions}a-b show examples of different realisations of such rare transitions for one choice of parameters. The works of \citet{VanKan2019a} and \citet{DeWit2022} have revealed that the waiting time that is spent until these sudden transitions from one state into the other commence is exponentially distributed, signifying that the transition process is memoryless. The typical mean waiting time $\tau_W$ can be obtained by defining representative thresholds in the large-scale energy and analysing the distribution of times $t_\textrm{b,d}$ after which these thresholds are crossed. Examples of the obtained empirical cumulative distributions are depicted in figure~\ref{fig:transitions}c, showing that it closely follows the aforementioned exponential distribution. We can then obtain $\tau_W$ by fitting the empirical cumulative distribution function (CDF) with a (shifted) exponential as
\begin{equation}
    \textrm{CDF}(t_\textrm{b,d})=1-\exp\left(-\frac{t_\textrm{b,d}-\tau_0}{\tau_W}\right).
\end{equation}
This process can be repeated to obtain the waiting time scales $\tau_W$ for build-up events and decay events at different values of $Q$, varying it across the full extent of the hysteretic transition. This results in a series of typical waiting times $\tau_W$ as a function of $Q$, which can in turn be repeated for different box sizes and $\Rey$.

\section{Time-scale statistics}\label{sec:results_discussion}     
\begin{figure}
    \centering
    \includegraphics[width=\linewidth]{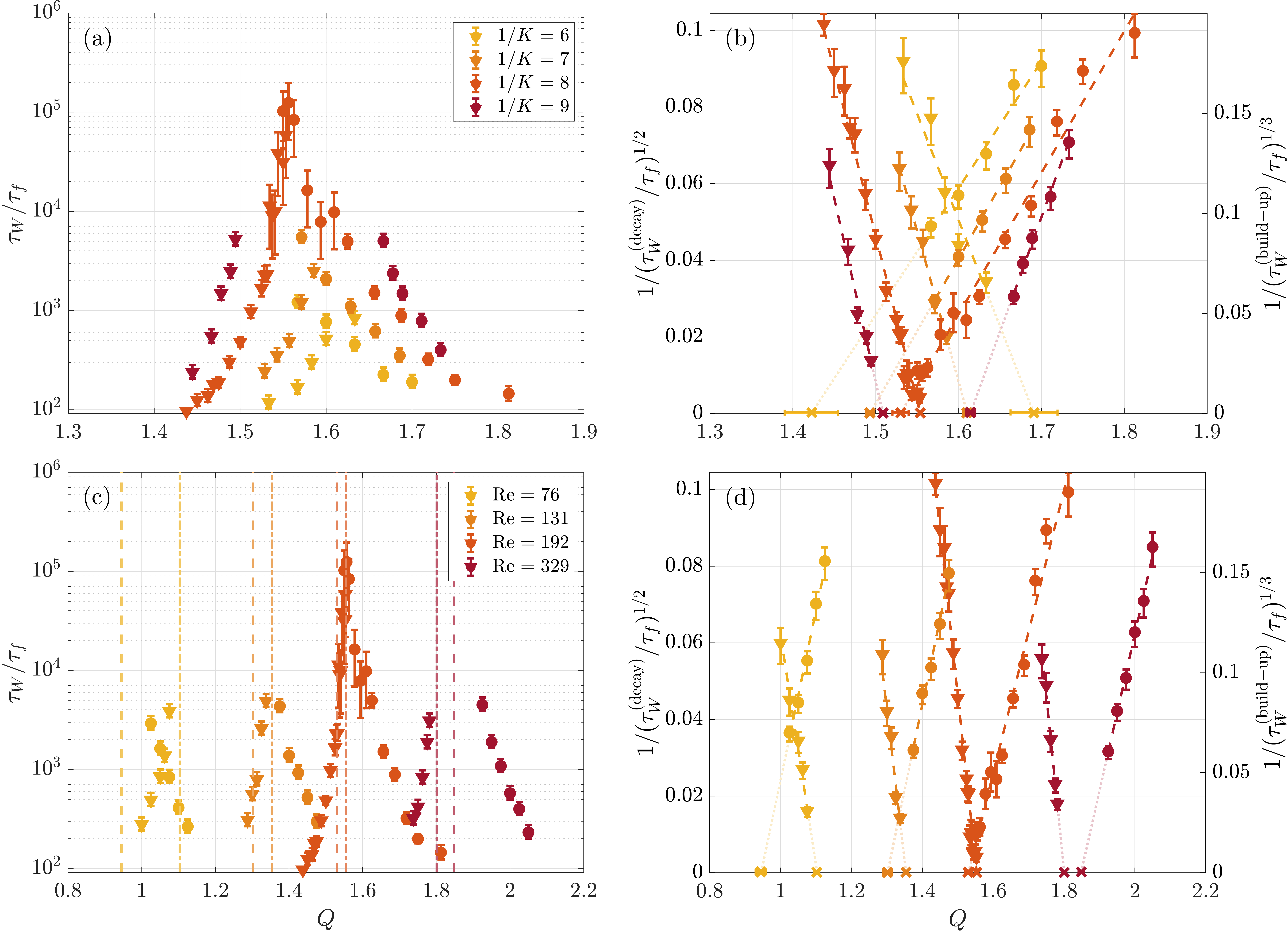}
    \caption{(a,c) Waiting times $\tau_W$ for build-up (circles) and decay (triangles) events and (b,d) their power-law transformation for varying box size $1/K$ at $\Rey=192$ (a,b) and varying $\Rey$ at $1/K=8$ (c,d). Crosses on the horizontal axis in (b,d) denote the estimates for the asymptotes $Q_0$. These asymptotes are also depicted in (c) by the vertical lines for build-up (dashed) and decay (dashed-dotted), but are omitted in (a) for readability.}
    \label{fig:tauw}
\end{figure}

The results for the series of transition time scales for varying box size and varying $\Rey$ are provided in figure~\ref{fig:tauw}(a,c). As the transition is approached, $\tau_W$ increases faster than exponentially \citep[see][]{VanKan2019a}. This implies that either $\tau_W$ increases in a non-diverging super-exponential fashion (e.g. $\tau_W\propto \exp[\exp(Q)]$, as is typical for certain transitions controlled by extreme events \citep{Goldenfeld2010,Nemoto2018,Nemoto2021,Gome2021}), or that it diverges at some critical value $Q_0$. To determine which of the two holds for the present system is beyond the scope of this work. We will thus assume that the latter case applies, although one could alternatively interpret $Q_0$ as the value at which super-exponential behaviour starts in the former case. To determine $Q_0$, we fit the transition time to a power-law divergence
\begin{equation}\label{eq:asymptote}
    \tau_W\propto\frac{1}{\left|Q-Q_0\right|^p}.
\end{equation}
By plotting $1/\tau_W^{1/p}$ as a function of $Q$, we can then obtain $Q_0$ from a linear fit to the data, see figure \ref{fig:tauw}(b,d). Empirically, we find that $p^{\textrm{(build-up)}}=3$ and $p^{\textrm{(decay)}}=2$ result in a satisfactory linearisation of our data. These asymptotes then predict the location in $Q$ where the transition time becomes infinite, such that we can say that beyond the asymptote, the transition cannot occur at any finite time scale.

Comparing the results at different box sizes and different $\Rey$, we find first of all that the transition is observed in a similar range of $Q$ for the varying box sizes, while it clearly shifts as $\Rey$ is varied, in agreement with the observations in \cite{VanKan2019}. More importantly, we observe that the branches of build-up and decay times move further apart as $\Rey$ and box size are increased: the build-up branch moves to larger $Q$, while the decay branch moves to (relatively) smaller $Q$. The branches cross for the runs with $\Rey\le 192$ or for $K\le 8$, such that at small box size and/or small $\Rey$, a \textit{bimodal} range of $Q$ exists for $Q_0^\textrm{(build-up)}<Q<Q_0^\textrm{(decay)}$ where the build-up and decay time scales are simultaneously finite (and in fact computationally accessible). Hence, in this range, the flow continually transitions back and forth between the 3D state and the condensate state. 

\begin{figure}
    \centering
    \includegraphics[width=\linewidth]{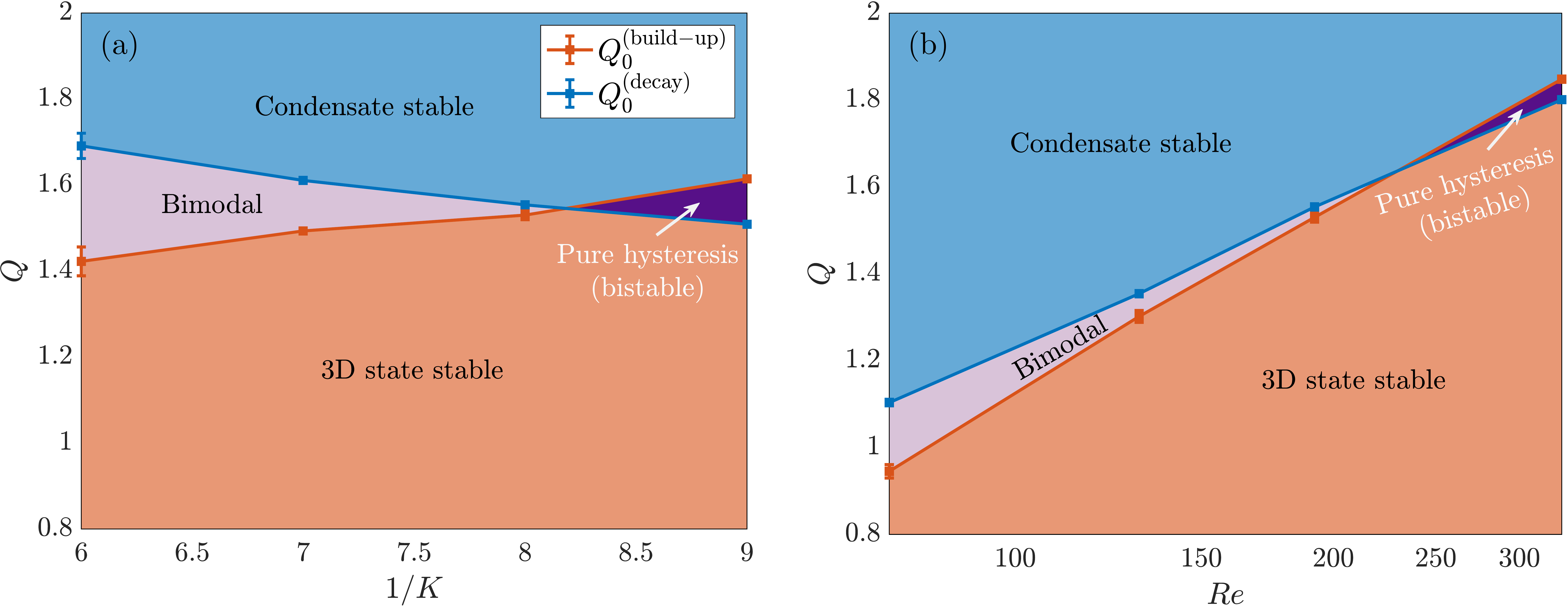}
    \caption{Phase diagrams of the condensate for (a) varying box size at $\Rey=192$ and (b) varying $\Rey$ at $1/K=8$. Red and blue lines denote the asymptotes of waiting times for build-up and decay events, respectively.}
    \label{fig:phase_diagrams}
\end{figure}

Conversely, for the largest box size and largest $\Rey$ that we consider, the branches of the build-up and decay branches never cross as the asymptotes reside on opposite ends. This indicates a profoundly different regime, where the decay times have diverged before the build-up times become finite, such that in the range $Q_0^\textrm{(decay)}<Q<Q_0^\textrm{(build-up)}$, both states are absolutely stable as no transition from one state into the other can occur in any finite time. This corresponds to a regime of \textit{pure hysteresis} in which the system is (absolutely) bistable. Indeed, it is this range that is arguably the most unambiguous time-scale-independent definition of the bistable range of the system. These results are summarised in figure~\ref{fig:phase_diagrams}. It is thus evident from our results that the bistable range grows as the box size and/or $\Rey$ are independently increased.

We argue that the strengthening of the bistability for large box sizes and $\Rey$ is intuitive from the increase of the condensate energy level (compared to the 3D-state energy) as $\Rey,1/K$ is increased, making it harder to jump from one state into the other. Our results thus indicate that bistability is not a finite-size, finite-$\Rey$ effect, and such states could possibly be found in the geophysical limit where both $\Rey$ and domain size are large.

\section{Conclusions and outlook}\label{sec:conclusions} 
In this work, we have demonstrated that the bistability observed in the transition to the quasi-2D condensate state persists as the box size and $\Rey$ are increased. By studying the time scales at which rare transitions from one state into the other occur, we quantified the precise extent of the bistable range. Fitting the mean time scales of these transitions with a diverging power law, we measured the locations of the asymptotes beyond which the transition is prohibited at any finite time scale. Since these asymptotes show a crossover as we vary the box size and/or $\Rey$, this predicts a profound regime change from a bimodal regime at small box size and/or $\Rey$ to a regime of pure hysteresis at large box size and/or $\Rey$, as summarised in figure~\ref{fig:phase_diagrams}. Since we find that the branches of time scales of build-up transitions into the condensate and decay transitions out of the condensate at both ends of the bistable range only separate further and further as the box size and/or $\Rey$ is increased, we conclude that this bistability is not a finite-size or finite-$\Rey$ effect, but that the bistable range grows as we progress towards the limit of increasing system size and/or $\Rey$.

We remark that the method proposed here for quantifying the precise extent of the bistable range in a hysteretic transition using the time scales of rare transitions is entirely general and a similar procedure can be followed in the context of any other fluctuating hysteretic dynamical system within or beyond fluid dynamics. However, we must note that the motivation of \eqref{eq:asymptote} is \textit{ad hoc} here. Although the agreement with our data as shown in \ref{fig:tauw} is satisfactory, a more fundamental physical motivation, supported by a larger dynamic range of parameters, would be needed to rigorously prove the validity of \eqref{eq:asymptote} as well as our choice of exponents. Indeed, although the waiting time increases faster than exponentially, the existence of an asymptote for $\tau_W$ in the first place is ultimately an assumption in itself and the possibility of the relation being any other super-exponential relation without divergence can in principle not be ruled out by numerical simulations alone. Nonetheless, while the existence of pure hysteresis certainly constrains the underlying physical mechanism of the transition, one may argue that it does not hold immediate implications in geophysical practice whether the time scales of transitions are strictly infinite or merely beyond any practical finite time scale. Moreover, we argue that the satisfactory agreement of \eqref{eq:asymptote} with our data in itself does convincingly prove our central result: that the bistable range is not an effect of finite box size or finite $\Rey$, but that it grows with increasing system size and/or $\Rey$. Indeed, this holds either in the strict terms of absolute bistability, or in terms of exceeding a certain finite super-exponential time scale.

While recent investigations have started to unveil different aspects of this peculiar type of transition between turbulent flow states, much of the underlying physical mechanism still remains poorly understood. In particular, which specific physical events trigger the flow to commence the transition -- for example, either a series of vortex merging events, or rare fluctuations directly at the largest scale -- remains an open question. Answering such questions would contribute greatly to our understanding of this flow phenomenon, and our work may act as a numerical inspiration as well as a quantitative benchmark to such theoretical studies. Specifically, understanding the physical mechanism behind the transition may motivate the theoretical validity of relation \eqref{eq:asymptote}, which we have motivated only empirically here.

The high computational demands of the presented analysis limit the range of accessible parameters in this work to rather moderate values at small grids. It is known from \cite{VanKan2019} that the observed increase of the critical values $Q_0$ with $\Rey$ in the examined range will eventually saturate at very large $\Rey$. The persistence of bistability at this asymptotic regime thus needs to be formally validated. Furthermore, we also limited ourselves to the simplest case of a thin layer forced by a 2D body force. More natural 3D forcing as well as effects such as rotation and stratification would also be needed to make contact with geophysical flows that would require a larger set of simulations to cover the high-dimensional parameter space, posing yet higher computational demands.

A promising solution may lie in the application of rare-event algorithms \citep{Cerou2007,Lestang2018}. Such algorithms are more efficient in probing rare transitions and have been successfully applied in various other flow contexts \citep{Gome2021,Bouchet2019,Rolland2018}. By progressing further towards the extreme geophysical conditions, we may for example investigate whether the growth of the bistable range of the condensate transition saturates at some point, which cannot be studied from the moderate parameters considered in our work. Finally, we remark that it also seems attractive now to study the condensate transition and its bistable behaviour from experiments in which more extreme parameters may be more easily accessible, or perhaps even from observations in real-world geophysical or astrophysical flows.

\backsection[Acknowledgments]{The authors thank H.J.H. Clercx for useful discussions and comments.}

\backsection[Funding]{This work was granted access to the HPC resources of MesoPSL financed by the Region Ile de France and the project Equip@Meso (reference ANR-10-EQPX-29-01) of the programme Investissements d'Avenir supervised by the Agence Nationale pour la Recherche and the HPC resources of GENCI-TGCC \& GENCI-CINES (Project No. A0090506421, A0110506421). This work has also been supported by the Agence nationale de la recherche (ANR DYSTURB project No. ANR-17-CE30-0004). AvK acknowledges support by Studienstiftung des deutschen Volkes and the National Science Foundation (grant DMS-2009563).}

\backsection[Declaration of Interests]{The authors report no conflict of interest.}

\bibliographystyle{jfm}
\bibliography{main}

\end{document}